\documentstyle[aps,prl,multicol,epsf]{revtex}

\begin{document}
\title{New mechanism for the enhancement of $sd$ dominance in interacting
boson models}
\author{J. Dukelsky$^{1}$ and S. Pittel$^{2}$}
\address{ $^1$ Instituto de Estructura de la Materia, Consejo
Superior de Investigaciones Cientificas, Serrano 123, 28006 Madrid,
Spain \\ 
$^2$Bartol Research Institute, University of Delaware,
Newark, Delaware 19716, USA }
\maketitle

\begin{abstract}
We introduce an exactly solvable model for interacting bosons that extend up to high spin
and interact through a repulsive pairing force. The model exhibits a phase
transition to a state with almost complete $sd$ dominance. The
repulsive pairing interaction that underlies the model has a natural microscopic
origin in the Pauli exclusion principle between contituent nucleons. As such, repulsive pairing between bosons seems to provide a new mechanism for the enhancement of $sd$ dominance, giving further support for the validity of the $sd$ Interacting Boson Model.  
 \end{abstract}

\begin{center}
\end{center}
\begin{multicols}{2}

The Interacting Boson Model (IBM)\cite{IBM} has been highly successful in describing and
correlating the collective properties of medium mass and heavy nuclei throughout the periodic table. The model, though
phenomenological in application, is deeply linked to the underlying nuclear shell model.
The $s$ and $d$ bosons of the model represent the lowest pair degrees of freedom of identical nucleons. The model has a $U(6)$ group structure and three posssible dynamical symmetry limits: 
\begin{equation}
U(6)\supset \left\{ 
\begin{array}{l}
U(5)\supset O(5)\supset O(3) \\ 
SU(3)\supset O(3) \\ 
O(6)\supset O(5)\supset O(3)
\end{array}
\right.  \label{U6}
\end{equation}
In each of the three limits, all of which have been realized to good approximation in nuclei, the hamiltonian can be expressed as a linear combination of the Casimir operators from the associated group chain. Most nuclei do not live at the symmetry limits, however. For a general nucleus, the hamiltonian can be expressed as a linear combination of the Casimir operators of all three group chains and the nucleus can be represented as a point inside a triangle with the three symmetry limits at the vertices \cite{casten}. 

The $U(5)$ chain was initially proposed to describe phenomenologically
vibrational nuclei. Soon after it was provided a microscopic interpretation using a
mapping procedure based on the Generalized Seniority approximation\cite{OAI}. 
The other two chains, describing axially deformed ($SU(3)$) and gamma-soft ($O(6))$ nuclei, are less well understood from a microscopic point of view. 

As noted earlier, the $s$ and $d$ bosons represent the lowest pair degrees of freedom of identical nucleons. There are of course other pairs as well, but they typically lie higher in energy. The key assumption of the IBM is that these other pair degrees of freedom can be ignored, except for their renormalized effects on operators in the $sd$ subspace. On the other hand, there has been no convincing microscopic demonstration of this effective decoupling except in vibrational nuclei.

In this letter, we point out a new mechanism for producing $sd$ dominance in boson models of nuclei. The mechanism has at its heart the interaction between alike bosons. We will show, by considering a new class of exactly-solvable boson
models, that the effects of higher-spin bosons are suppressed because of the repulsion at short range of the boson-boson interaction. Furthermore, we claim that a short-range repulsive boson-boson interaction is ubiquitous as it is a direct reflection of the Pauli principle at the underlying nucleon level.

The models we consider involve even angular momentum bosons extending up to fairly high spin and interacting via a repulsive boson pairing interaction. As we will demonstrate, they can be solved exactly using a method introduced by Richardson in the 1960s\cite{Richardson} and recently generalized\cite{Schuck} to confined boson systems. The models are closely related to an exactly solvable model proposed to describe the transition between the $O(6)$ and $U(5)$ symmetry limits of the IBM\cite{Draayer}. The primary difference is that the model of Ref. 6 was restricted to $s$ and $d$ bosons, whereas ours accomodate higher-spin bosons as well. 

The models we study are based on an $SU(1,1)$ group algebra, for which the generators are

\[
K_{l}^{0}=\frac{1}{2}\sum_{m}\left( a_{lm}^{\dagger }a_{lm}+\frac{1}{2}%
\right) =\frac{1}{2}\widehat{n}_{l}+\frac{1}{4}\Omega _{l} ~,
\]

\begin{equation}
K_{l}^{+}=\frac{1}{2}\sum_{m}\left( -\right) ^{l-m}a_{lm}^{\dagger
}a_{l-m}^{\dagger }=\frac{1}{2}A_{l}^{\dagger } ~,
\label{gen}
\end{equation}
where the operator $a_{lm}^{\dagger}\left( a_{lm}\right) $ creates
(destroys) a boson with angular momentum $lm$, $\,$and $\Omega _{l}=2l+1$.
We have also defined operators $\widehat{n}_{l}$ and $A_{l}^{\dagger }$
which can be used instead of the generators $K$.

In terms of the $SU(1,1)$ generators (\ref{gen}), the most general hermitian
one- and two-body operator that preserves the number of bosons (commutes with
the number operator) can be written as

\begin{eqnarray}
R_{l}=K_{l}^{0}+2g\sum_{l^{\prime }\neq l}
&&\left\{ 
\frac{X_{ll^{\prime }}}{2}
\left[ K_{l}^{+}K_{l^{\prime }}^{-}+K_{l}^{-}K_{l^{\prime }}^{+}\right] 
\right.
\nonumber
\\
&&~
\left.
-Y_{ll^{\prime }}K_{l}^{0}K_{l^{\prime }}^{0}\right\} ~, \label{R}
\end{eqnarray}
where the matrices $X$ and $Y$ are at this point arbitrary.

We will consider the dynamics of an $N$ boson system in a Hilbert space that is
classified by the product group space

\begin{equation}
SU(1,1)_{s}\times SU(1,1)_{d}\times \cdots SU(1,1)_{L}  ~, \label{space}
\end{equation}
where $L$ is the maximum (even) angular momentum permitted. In this space a model is
fully integrable if the set of $P=L/2+1$ global operators $R$~commute
with one another, viz. 

\begin{equation}
\left[ R_{l},R_{l^{\prime }}\right] =0  ~~. \label{commu}
\end{equation}

We will concentrate in this work on a specific class of integrable models. These models correspond 
to the solution of (\ref{commu}) for which 
$X_{ll^{\prime }}=Y_{ll^{\prime }}=\frac{1}{\eta _{l}-\eta _{l^{\prime }}}~$,
where the $\eta ^{\prime }s$ are a set of $P$ arbitrary real numbers. The $R$
operators of the models can be expressed more compactly as

\begin{equation}
R_{l}\left( g,\eta \right) =K_{l}^{0}+2g\sum_{l^{\prime }\neq l}\frac{%
K_{l}\cdot K_{l^{\prime }}}{\eta _{l}-\eta _{l^{\prime }}}  ~.
\label{integ}
\end{equation}

It is straightforward to show that the specific linear combination $H_{PM}=2\sum_{l}%
\varepsilon _{l}R_{l}\left( g,\varepsilon \right) $ of the $R$ operators (\ref{integ}) is precisely
the standard pairing model (PM) hamiltonian

\begin{equation}
H_{PM}=\sum_{l}\varepsilon _{l}\widehat{n}_{l}+\frac{g}{2}\sum_{ll^{\prime
}}A_{l}^{\dagger }A_{l^{\prime }}+c  ~,
\label{PM}
\end{equation}
where $c$ is a constant.

We now return to the set of $R$ operators (\ref{integ}) and search for their mutual eigenvectors.
To do this, we follow the procedure introduced by Richardson for the pairing model hamiltonian (\ref{PM}).
Namely, we consider the ansatz

\begin{equation}
\left| \Psi \right\rangle =\prod_{\alpha =1}^{M}B_{\alpha }^{\dagger }\left|
\nu \right\rangle ~,\qquad B_{\alpha }^{\dagger }=\sum_{l}\frac{1}{2\eta
_{l}-e_{\alpha }}K_{l}^{+}  ~, \label{psi}
\end{equation}
where $M=(N-\nu)/2$ is the number of boson pairs coupled to zero angular momentum, and $\left| \nu \right\rangle \equiv \left| \nu _{s},\nu _{d},\cdots ,\nu
_{L}\right\rangle $ is a state with $\nu _{l}$ 
unpaired bosons with angular momentum $l$, defined by

\[
K_{l}^{0}\left| \nu \right\rangle =k_{l}\left| \nu \right\rangle
~,~K_{l}^{-}\left| \nu \right\rangle =0 ~,
\]
with $k_{l}=\nu _{l}/2+\Omega _{l}/4$ and $\nu = \sum_{l} \nu_l$.

Acting with the $R$ operators on the state (\ref{psi}) and considering the conditions that
must be satisfied to
fulfill the eigenvalue equation $R_{l}\left| \Psi \right\rangle =r_{l}\left|
\Psi \right\rangle $, we obtain a set of coupled equations for the unknown
pair energies $e_{\alpha }$,

\begin{equation}
1+4 g\sum_{l}\frac{k _{l}}{2\eta _{l}-e_{\nu }}+4g\sum_{\mu \neq \nu }%
\frac{1}{e_{\mu }-e_{\nu }}=0 ~. \label{richar}
\end{equation}

This set of equations was first derived by Richardson \cite{Richardson} when considering the eigenstates of the standard pairing model hamiltonian ($\eta_l = \varepsilon_l$). He showed that there are as many solutions as states in the Hilbert space. Thus, the
equations (\ref{richar}) define a complete set of eigenstates of the form (\ref{psi}).

The pair energies 
$e_{\mu }$ are always real for boson systems. The different states are classified by their configuration 
in the (weak-coupling) limit in which the pair energies $e_{\mu }$ tend to their unperturbed values $2 \eta _{l}$. In particular, the ground state (GS) corresponds to all pair energies in the interval $2 \eta_{s}< e_{\mu } < 2\eta_{d}$. Excited states can be obtained either by breaking pairs or by promoting pairs to higher-energy intervals. 
As an example, the first excited state with the same seniority (i.e., the same $\nu$) as the GS will have $M-1$ pair energies in the interval $2 \eta_{s}< e_{\mu } <2 \eta_{d}$ and one pair energy in the interval $2 \eta_{d}< e_{\mu } < 2 \eta_{g}$.           
Coming back to the set of equations (\ref{richar}), we now see that it is a general condition for the common eigenstates of the mutually-commuting $R$ operators. Thus, it can be used for finding the eigenstates of {\em any} linear combination of the $R$ operators, not just the linear combination that corresponds to the pure pairing hamiltonian.
Once we have solved (\ref{richar}) for the pair energies, we can readily obtain the eigenvalues of the corresponding $R$ operators  from

\[
r_{l}=k_{l}\left[ 1-4g\sum_{\nu }\frac{1}{2\eta _{l}-e_{\nu }}%
-2g\sum_{l^{\prime }\left( \neq l\right) }\frac{k_{l^{\prime }}}{\eta
_{l}-\eta _{l^{\prime }}}\right] ~.
\]

A {\em general} pairing hamiltonian can be expressed as a linear combination
of the $R$ operators as

\begin{equation}
H=2\sum_{l}\varepsilon _{l}R_{l}\left( g,\eta \right)   \label{HP} ~.
\end{equation}
Replacing the $K_{l}$ operators in the $R_{l}$ operators appearing in (\ref{HP}) by the pair operators $A_{l}$ and
the number operators $\widehat{n}_{l},$ and adding and substracting the $S(1,1)_{l}$
Casimir operator, we obtain 
\begin{equation}
H=A+\sum_{l}\epsilon _{l}\widehat{n}_{l}+\sum_{ll^{\prime }}V_{ll^{\prime }}\left[
A_{l}^{\dagger }A_{l^{\prime }}-\widehat{n}_{l}\widehat{n}_{l^{\prime }}\right]  ~,
\label{HPP}
\end{equation}
where 

\[
A=\frac{1}{2}\sum_{l}\epsilon _{l}\Omega _{l}+\frac{1}{4}%
\sum_{l}V_{ll}\Omega _{l}^{2}-\frac{1}{4}\sum_{ll^{\prime }}V_{ll^{\prime
}}\Omega _{l}\Omega _{l^{\prime }} ~,
\]

\[
\epsilon _{l}=\varepsilon
_{l}+2V_{ll}-\sum_{l^{\prime }}V_{ll^{\prime }}\Omega _{l^{\prime }} ~,
V_{ll^{\prime }}=\frac{g}{2}\frac{\varepsilon _{l}-\varepsilon _{l^{\prime }}%
}{\eta _{l}-\eta _{l^{\prime }}} ~.
\]

As noted earlier, for $\varepsilon_l=\eta_l$, we recover the standard PM  hamiltonian (\ref{PM}). Restricting (\ref{HPP}) to $l=0$ and $2$ bosons only, we obtain a hamiltonian very similar to that of Ref. 6, the only difference being that our hamiltonain has an additional monopole-monopole interaction term that can be traced to the term involving
$Y_{ll^{\prime}}$ in (\ref{R}).  

\begin{figure}
\hspace{0.1cm}
\epsfysize=7cm
\epsfxsize=8cm
\epsffile{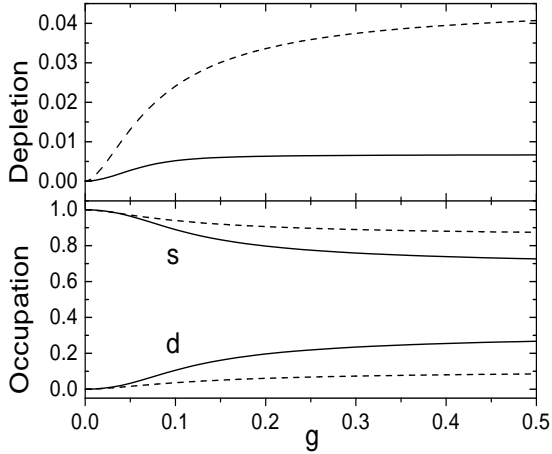}
\narrowtext

\caption{Occupation probabilities for the ground state of a system of $M=5$ boson pairs and maximum
angular momentum $L=12$ as a function of the interaction strength $g$. The
upper graph shows the sum of occupation probabilities (depletion) for 
high-spin bosons ($l>2$) while the lower graph gives the occupation probabilities
for $s$ and $d$ bosons.}
\label{fig1}
\end{figure}

An important feature of our exactly solvable and integrable hamiltonian (\ref{HPP})
is that it is not restricted to an $sd$ subspace but also permits
higher angular momentum bosons. We will now make use of this property to study
the effect of the inclusion of high-spin bosons on the $sd$ subspace
dynamics.

Since there is virtually no information available on the coupling of the usual IBM space to 
higher-spin bosons, we will consider two sample scenarios. In the first (denoted I), we
assume a constant pairing interaction, as arises for $\eta _{l}=\varepsilon _{l}$. We also assume
linear single-boson
energies ($\varepsilon _{l}=l$). 

As is well known, a pure pairing interaction produces
enhanced pair scattering to high-spin states. This is presumably a consequence of the extreme short-range nature of a pairing force, and is reflected in the fact that the {\em effective} pairing matrix element that connects a level with angular momentum $l$ to one with angular momentum $l^{\prime}$ is $V_{ll^{\prime}}^{eff}= V_{ll^{\prime}} \sqrt{(2l+1)(2l^{\prime}+1)}~.$
This anomalous feature of a boson pairing force bears some similarity to the well-known pathology of a delta interaction in fermion theories. There, the unphysical behavior is typically overcome by introducing a cutoff in energy.  In our boson context, we can likewise modify the interaction to get rid of its undesirable high-$l$ behavior. In particular, if we select the $\eta ^{\prime}s $  according to $\eta _{l}=l^{2}$, the effective pairing matrix elements would
scale as
\[
V_{ll^{\prime}}^{eff}  \sim \frac{g\sqrt{\left( 2l+1\right) \left( 2l^{\prime }+1\right) }}{%
2\left( l+l^{\prime }\right) }\sim 1 ~ , 
\]
and not produce enhanced scattering to high-spin pairs. In the second scenario (denoted II), we will make this assumption for the $\eta$'s and again assume linear single-boson energies ($\varepsilon_l=l$). While we feel that model II is more realistic than model I, we will for completeness present results for both.

For both models I and II, as well as for any other general pairing model, the occupation probabilities can be readily obtained from the $R$ operators according to

\[
\left\langle \widehat{n}_{l}\right\rangle =-\frac{\Omega _{l}}{2}%
+2\left\langle R_{l}\right\rangle -2g\left\langle \frac{\partial R_{l}}{%
\partial g}\right\rangle ~.
\]

Using the Hellmann-Feynman theorem, we can replace the $R$ operators by their eigenvalues to obtain an equation for the occupation numbers in terms of the derivatives of the pair energies with respect to the pairing strength $g$. Furthermore, taking the derivative of the Richardson equations (\ref{richar}) with respect to $g$ leads to a set of linear equations for the pair energy derivatives.   
In Fig. 1, we show the occupation probabilities for the GS of a system of $10$ bosons
and an angular momentum cutoff of $L=12$ as a function of $g$. In the lower
part of the figure we show the occupation probabilities for $s$ and $d$
bosons, while in the upper part we show the summed occupation probabilities
of those bosons with $l>2$, which we call the depletion of the IBM space. In
both parts of the figure, the dashed line refers to model I and the solid
line to model II.

It is apparent from the figure that the class of models we have considered displays a quantum
phase transition from an $s\,$boson condensate (spherical) to a mixed
state of $s$ and $d$ bosons (``deformed"). The transition is softened by the
finite number of bosons so that there is still some occupation probability of
high-spin bosons. Two important features should be noted in these results. One is that the
depletion increases in the spherical phase up to the phase transition
and then stays almost constant in the deformed phase. The other is
related to the choice of the model interaction. In the first scenario with
the PM hamiltonian and linear single-boson energies the depletion is small but non-negligible; 
in the more physical second scenario the depletion is negligible and $sd$ dominance is almost complete.

The fact that these models all lead to a very small non-$sd$ content in their ground states is the key conclusion of this paper. It suggests that repulsive pairing is indeed a robust mechanism for reinforcing $sd$ dominance in boson models of nuclei.

The above analysis was for a system of 10 bosons. An important issue not addressed by those results concerns the dependence on the number of bosons. How many bosons are needed for the repulsive pairing interaction to be effective in suppressing the high-spin content of the low-lying states?

This question is addressed in Fig. 2, where we show the $s$ and $d$ occupations and the IBM depletion for the GS as a function of the number of boson pairs $M$ for the specific value of $g=0.5$. As is evident from Fig. 1, this choice of the pairing strength places the system within the deformed phase, where there is mixing of different bosons in the ground state.

\begin{figure}
\hspace{0.1cm}
\epsfysize=7cm
\epsfxsize=8cm
\epsffile{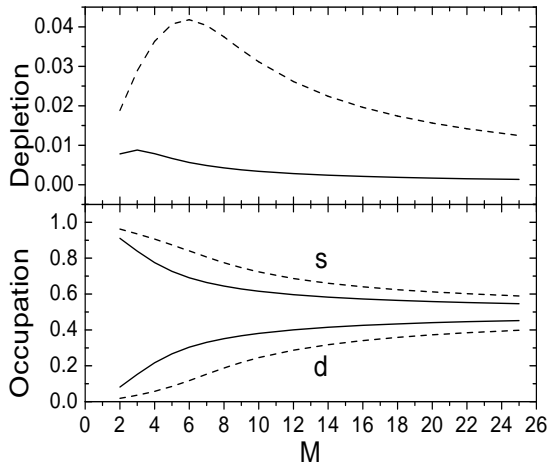}
\narrowtext

\caption{GS occupation probabilities for a system of bosons and maximum angular
momentum $L=12$ and pairing strength $g=0.5$ as a function of the number of
boson pairs $M$. The upper graph shows the sum of occupation probabilities
for high-spin bosons ($l>2$) while the lower graph gives the occupation
probabilities for $s$ and $d$ bosons.}
\label{fig2}
\end{figure}

The first point to note is that for small enough boson number, the depletion from the $sd$ subspace actually increases with increasing boson number. What is important, however, is that for both hamiltonians a critical boson number is soon reached after which the depletion decreases and $sd$ dominance gradually becomes better and better. The precise number of bosons required for this critical behavior to set in depends on the hamiltonian. It is 12 for the standard pairing interaction and 6 for the more realistic modified pairing interaction.

It is interesting to ask why in the presence of a repulsive pairing interaction there is a phase transition to a state of essentially pure $sd$ content. 

As mentioned before, the GS solution of (\ref{richar}) for repulsive pairing corresponds 
to all pair energies in the interval $2 \eta_s < e_\mu < 2 \eta_d$. This means that 
the pair operators in the Richardson ansatz (\ref{psi}) have one phase for $s$ boson
pairs and the opposite phase for {\em all} other boson pairs. It is only when this set of phase relations is satisfied that the system can gain in energy from a repulsive pairing interaction. On the other hand, for two given boson degrees of freedom, the only way they can take advantage of a repulsive pairing force is by having opposite phases. Putting these two facts together, we see that only one of the non-$s$ boson degrees of freedom can correlate with the $s$ boson and produce a gain in energy in the presence of a repulsive pairing interaction. Clearly, this will be the $d$ boson since it has the lowest energy. The picture that emerges is that the $d$ boson mixes with the $s$ boson once the pairing interaction is strong enough to overcome the difference in single-boson energies, thus explaining the phase transition from $SU(5)$ to $O(6)$ in the IBM as a function of the pairing strength. 
Higher-spin bosons do not lead to significant further correlations, since they cannot have opposite phases to {\em both} the $s$ and $d$ bosons that make up the $O(6)$ wave function.

At this point it is worth expanding briefly on the comment made earlier that a short-range repulsive interaction between composite objects is a natural consequence of the Pauli principle at the constituent level. That this is the case is well known both in molecular physics, where the short-range repulsion between the two hydrogen atoms in an $H_2$ molecule is the result of electron exchange, and in nuclear physics, where the nuclear force has a strong short-range repulsion due to quark exchange. It also arises naturally when boson mapping methods are applied to systems with spatial two-body correlations\cite{Delta}. 

In this work, we have discussed a class of boson models involving a generalized repulsive  
pairing interaction for bosons with arbitrary even angular momenta. We have shown that these models are exactly solvable using a method originally developed by Richardson for more traditional pairing hamiltonians. We have applied this method to a system in which the boson degrees of freedom extend to a cutoff angular momentum significantly larger than 2. Nevertheless, we find that the ground state of the system is almost completely dominated by $s$ and $d$ bosons, the effect of higher-spin bosons being strongly suppressed by the repulsive pairing interaction.
Considering that a repulsive pairing interaction between bosons is a natural consequence of the Pauli principle at the nucleon level, these results provide further microscopic support for the validity of the interacting boson model in nuclear structure.

This work was supported in part by the National Science Foundation
under grant \# PHY-9970749, by the Spanish DGI under grant
BFM2000-1320-C02-02, and by NATO under grant PST.CLG.977000.

\end{multicols}

\end{document}